\documentclass[12pt]{article}
\def\be{\begin{equation}}
\def\ee{\end{equation}}
\def\bea{\begin{eqnarray}}
\def\eea{\end{eqnarray}}
\usepackage{graphicx}

\catcode`\@=11
\def\lsim{\mathrel{\mathpalette\@versim<}}
\def\gsim{\mathrel{\mathpalette\@versim>}}
\def\@versim#1#2{\vcenter{\offinterlineskip
\ialign{$\m@th#1\hfil##\hfil$\crcr#2\crcr\sim\crcr } }}
\catcode`\@=12

\catcode`@=12
\begin{document}
\thispagestyle{empty}
\begin{flushright}
UCRHEP-T523\\
September 2012\
\end{flushright}
\vspace{1.0in}
\begin{center}
{\Large \bf Self-Organizing Neutrino Mixing Matrix\\}
\vspace{0.5in}
{\bf Ernest Ma\\}
\vspace{0.2in}
{\sl Department of Physics and Astronomy, University of
California,\\
Riverside, California 92521, USA\\}
\vspace{0.1in}
{\sl Centre for Theoretical Physics (CFTP), Technical University of 
Lisbon,\\
1049-001 Lisbon, Portugal\\}
\end{center}
\vspace{01.5in}
\begin{abstract}\
A new and novel idea for a predictive neutrino mass matrix is 
presented, using the non-Abelian discrete symmetry $A_4$ and the 
seesaw mechanism with only two heavy neutral fermion singlets. 
Given the components of the one necessarily massless neutrino 
eigenstate, the other two massive states are automatically generated. 
A realistic example is discussed with predictions of a normal 
hierarchy of neutrino masses and maximal ${CP}$ violation.
\end{abstract}

\newpage
\baselineskip 24pt

To understand the observed neutrino mixing pattern in terms of a 
symmetry, the charged-lepton mass matrix and the neutrino mass 
matrix must be considered at the same time.  Given that 
$m_{e,\mu,\tau}$ are all different, it is by no means trivial to 
find a symmetry which predicts a leptonic mixing matrix as the 
mismatch between the unitary matrices which diagonalize the 
respective mass matrices in the two different sectors.  This 
was successfully done using the non-Abelian discrete symmetry 
$A_4$~\cite{mr01,m02,bmv03} and applied~\cite{m04} to the case 
of tribimaximal mixing.   Whereas the specific prediction of 
$\theta_{13} = 0$ is now refuted by data~\cite{daya12,reno12}. 
it does not mean that $A_4$ itself is not valid, only those 
additional assumptions beyond $A_4$ which are used to enforce the 
tribimaximal hypothesis.  Two variations~\cite{mw11,im12} of 
the original $A_4$ model~\cite{m04} are in fact completely 
consistent with $\sin^2 2\theta_{13} = 0.1$.

In this paper, an entirely different application of $A_4$ is 
presented for a predictive neutrino mass matrix.  It is based on 
an earlier proposal~\cite{m05} which works very well if 
$\sin^2 2\theta_{13}$ is small~\cite{lk07} but not with present 
data~\cite{daya12,reno12}.  The new and novel idea is to combine 
the $A_4$ texture with the seesaw mechanism using only two heavy 
neutral fermion singlets.  As a result, a massless neutrino 
eigenstate must appear.  If it is identified with $\nu_1$, then 
$\nu_2$ and $\nu_3$ are generated with $m_2 = \sqrt{\Delta m^2_{21}}$ 
and $m_3 = \sqrt{\Delta m^2_{31}}$.  The tribimaximal case may in 
fact be derived this way in a certain symmetry limit.  Here it will 
be shown how a realistic pattern of masses and angles emerges, 
with predictions of the Dirac phase $\delta_{CP}$ for leptonic 
$CP$ violation and the effective mass $m_{ee}$ in neutrinoless 
double beta decay.

Before showing how $A_4$ allows this to happen, consider the 
end result, i.e.
\begin{equation}
{\cal M}_\nu = \pmatrix{(2A+2B)u_1^2 & (-A-B+iC)u_1u_2 & 
(-A-B-iC)u_1u_3 \cr (-A-B+iC)u_1u_2 & (2A-B-iC)u_2^2 & 
(-A+2B)u_2u_3 \cr (-A-B-iC)u_1u_3 & (-A+2B)u_2u_3 & 
(2A-B+iC)u_3^2}.
\end{equation}
Note that in this basis, the charged-lepton mass matrix is diagonal, 
which is not an assumption but a consequence of the $A_4$ symmetry.   
It is clear from the above that there is one 
massless eigenstate, i.e.
\begin{equation}
\nu_1 = (u_1^{-1}, u_2^{-1}, u_3^{-1})/\sqrt{u_1^{-2}+ u_2^{-2}+
 u_3^{-2}} 
\end{equation}
for any $A,B,C$.  Let $\nu_{1,2,3}$ be defined by the tribimaximal 
basis, i.e. 
\begin{equation}
\pmatrix{\nu_1 \cr \nu_2 \cr \nu_3} = \pmatrix{\sqrt{2/3} & 
-\sqrt{1/6} & -\sqrt{1/6} \cr 1/\sqrt{3} & 1/\sqrt{3} & 1/\sqrt{3} 
\cr 0 & -1/\sqrt{2} & 1/\sqrt{2}} \pmatrix{\nu_e \cr \nu_\mu \cr 
\nu_\tau},
\end{equation}
then for $u_1 = 1/2$, $u_2 = u_3 = -1$,
\begin{equation}
{\cal M}_\nu^{(1,2,3)} = \pmatrix{0 & 0 & 0 \cr 0 & 3(B+A)/2 & 
i \sqrt{3/2}  C \cr 0 & i \sqrt{3/2}  C & -3(B-A)}.
\end{equation}
This shows that for $C=0$, tribimaximal mixing is obtained with
\begin{equation}
m_1=0, ~~~ m_2=3(B+A)/2, ~~~ m_3=-3(B-A).
\end{equation}
However, since $C$ is in general not zero, deviation from 
tribimaximal mixing will occur, as shown below.

The form of the neutrino mass matrix of Eq.~(1) is diagonalized 
by the unitary matrix $U$, i.e.
\begin{equation}
U {\cal M}_\nu U^T = \pmatrix{0 & 0 & 0 \cr 0 & m_2 & 0 \cr 0 & 0 
& m_3},
\end{equation}
where 
\begin{eqnarray}
\nu_1 &=& \left( \sqrt{2 \over 3}, -{1 \over \sqrt{6}}, 
-{1 \over \sqrt{6}} \right), \\
\nu_2 &=& {1 \over \sqrt{1+3 \zeta^2}} \left( {1 \over \sqrt{3}}, 
{1 \over \sqrt{3}} + i \sqrt{3 \over 2}  \zeta, 
{1 \over \sqrt{3}} - i \sqrt{3 \over 2}  \zeta \right), \\
\nu_3 &=& {1 \over \sqrt{1+3 \zeta^2}} \left( -i \zeta, 
-{1 \over \sqrt{2}} - i \zeta, 
{1 \over \sqrt{2}} - i \zeta \right).
\end{eqnarray}
This solution is obtained with
\begin{eqnarray}
B+A &=& {2 \over 1+3\zeta^2} \left( {m_2 \over 3} - \zeta^2 m_3 
\right), \\ 
B-A &=& - {1 \over 1+3\zeta^2} \left( {m_3 \over 3} - \zeta^2 m_2 
\right), \\
C &=& - {\sqrt{2} \zeta \over 1+3\zeta^2} (m_3 + m_2).
\end{eqnarray}

Using Eqs.~(7),(8),(9), the mixing angles in the conventional 
definition are given by
\begin{equation}
\sin \theta_{13} = {\zeta \over \sqrt{1+3\zeta^2}}, ~~~ 
\sin \theta_{12} = {1 \over \sqrt{3} \sqrt{1+2\zeta^2}}, ~~~ 
\sin \theta_{23} = -{1 \over \sqrt{2}}. 
\end{equation}
As for $CP$ violation, using the Jarlskog invariant, it is 
easily shown that
\begin{equation}
\sin \delta_{CP} = 1,
\end{equation}
i.e. maximal $CP$ violation.  Note that Eqs.~(10),(11),(12) allow 
complex values of $m_2$ and $m_3$, but Eqs.~(7),(8),(9) remain 
the same, and so thus Eqs.~(13) and (14).  In other words, this model's 
three mixing angles and one Dirac phase are independent of the Majorana 
phases of $\nu_{2,3}$.

Using the  experimental 
constraints~\cite{pdg12}
\begin{eqnarray}
|m_2|^2 &=& 7.50 \pm 0.20 \times 10^{-5}~{\rm eV}^2, \\
|m_3|^2 &=& 2.32 + 0.12 (- 0.08) \times 10^{-3}~{\rm eV}^2,
\end{eqnarray}
and assuming $m_{2,3}$ to be real, the two cases of $m_2 = \pm 0.00866$ eV 
(with $m_3 = 0.04817$ eV) are considered, as well as $\zeta = 0.165$ from 
$\sin^2 2 \theta_{13} = 0.098$.  The parameter values of this model are 
then determined to be
\begin{eqnarray}
&& A = 0.08769~{\rm eV}, ~~~ B = -0.00586~{\rm eV}, ~~~ 
C = -0.01226~{\rm eV}, \\  
&& A = 0.00365~{\rm eV}, ~~~ B = -0.01141~{\rm eV}, ~~~ 
C = -0.00852~{\rm eV},
\end{eqnarray}
respectively.  The effective neutrino mass in neutrinoless double 
beta decay is $m_{ee} = |A+B|/2 = 0.04$ or 0.004 eV.  They represent the 
maximum and minimum values of $m_{ee}$ in the presence of arbitrary 
Majorana phases.  Note also that $\theta_{13}$ is related to $\theta_{12}$ by
\begin{equation}
\tan^2 \theta_{12} = {1-3\sin^2 \theta_{13} \over 2} < 1/2.
\end{equation}
This is a generic consequence of any model which has $\nu_1 \sim 
(2,-1,-1)$ and is favored by data.  In another class of models 
where $\nu_2 \sim (1,1,1)$, the relationship becomes
\begin{equation}
\tan^2 \theta_{12} = {1 \over 2-3\sin^2 \theta_{13}} > 1/2,
\end{equation}
which is disfavored by data.

It has been shown that the neutrino mass matrix of Eq.~(1) allows 
it to generate $\nu_{2,3}$ once the massless state 
$\nu_1 \sim (u_1^{-1},u_2^{-1},u_3^{-1})$ is decided.  It has 
the simple and verifiable predictions of Eqs.~(13) and (14), if 
$\nu_1 \sim (2,-1,-1)$. 
To derive Eq.~(1), the symmetry $A_4$ is used following 
Ref.~\cite{m05}.  The lepton and Higgs representations are 
listed in Table 1.
\begin{table}[htb]
\begin{center}
\begin{tabular}{|c|c|c|c|c|}
\hline
Particle & $SU(2)_L \times U(1)_Y$ & $A_4$ & $Z_2^{(1)}$ & 
$Z_2^{(2)}$ \\ 
\hline
$(\nu,l)_{1,2,3}$ & (2,--1/2) & 3 & + & + \\ 
$l^c_{1,2,3}$ & (1,1) & 3 & -- & + \\ 
$N^c_{2,3}$ & (1,0) & $1'$, $1''$ & + & -- \\ 
\hline
$(\phi^0,\phi^-)_{1,2,3}$ & (2,--1/2) & 1, $1'$, $1''$ & -- & + \\ 
$(\eta^+,\eta^0)_{1,2,3}$ & (2,1/2) & 3 & + & -- \\
\hline
\end{tabular}
\caption{Particle content of proposed $A_4$ model of neutrino mass.}
\end{center}
\end{table}
The important departure from Ref.~\cite{m05} is that $N^c_1 \sim 1$ 
is now missing.  The two $Z_2$ symmetries are used to distinguish 
the two different sets of Higgs doublets.  Because of the $A_4$ 
multiplication rules~\cite{mr01}, the charged-lepton mass matrix 
is diagonal with
\begin{equation}
\pmatrix{m_e \cr m_\mu \cr m_\tau} = \pmatrix{1 & 1 & 1 \cr 1 & 
\omega & \omega^2 \cr 1 & \omega^2 & \omega} \pmatrix{h_1 v_1 \cr 
h_2 v_2 \cr h_3 v_3},
\end{equation}
where $\omega = \exp (2 \pi i/3) = -1/2 + i \sqrt{3}/2$ and 
$v_{1,2,3}$ are the vacuum expectation values of $\phi^0_{1,2,3}$. 
The Dirac mass matrix linking $\nu_{e,\mu,\tau}$ to $N^c_{2,3}$ 
is now
\begin{equation}
{\cal M}_D = \pmatrix{f_2 u_1 & f_3 u_1 \cr f_2 \omega u_2 & f_3 
\omega^2 u_2 \cr f_2 \omega^2 u_3 & f_3 \omega u_3},
\end{equation}
where $u_{1,2,3}$ are the vacuum expectation values of 
$\eta^0_{1,2,3}$. The most general Majorana mass matrix for 
$N^c_{2,3}$ is given by
\begin{equation}
{\cal M}_N = \pmatrix{M_2 & M_1 \cr M_1 & M_3}.
\end{equation}
Note that $M_1$ is an invariant mass under $A_4$, but $M_{2,3}$ 
are soft terms which break $A_4$.  In this model, $A_4$ is broken 
spontaneously by $v_{1,2,3}$ and $u_{1,2,3}$ as well as softly in 
the complete Lagrangian.  After inverting ${\cal M}_N$ and using 
the seesaw formula ${\cal M}_\nu = -{\cal M}_D ({\cal M}_N)^{-1} 
{\cal M}_D^T$, Eq.~(1) is obtained with
\begin{equation}
A = -{f_2 f_3 M_1 \over M_1^2 - M_2 M_3}, ~~~ 
B = {f_2^2 M_3 + f_3^2 M_2 \over 2(M_1^2 - M_2 M_3)}, ~~~ 
C = {f_2^2 M_3 - f_3^2 M_2 \over 2(M_1^2 - M_2 M_3)}.
\end{equation}
The next step is to choose $\nu_1 \sim (u_1^{-1},u_2^{-1},u_3^{-1})$.  
Since $\nu_1$ is guaranteed to be massless, Eq.~(1) is reduced to 
a $2 \times 2$ mass matrix in the basis $\nu'_2 \sim [-u_1(u_2^{-2}+u_3^{-2}), 
u_2^{-1},u_3^{-1}]$ and $\nu'_3 \sim (0,u_2,-u_3)$.  Diagonalizing this then 
yields the two mass eigenstates $\nu_{2,3}$ with $m_{2,3}$ and the 
corresponding mixing angle and Dirac phase.  The special case of 
$\nu_1 \sim (2,-1,-1)$ has been studied in this paper, but the method 
may be adapted to any $\nu_1$.

In conclusion, a remarkable form of the neutrino mass matrix has been 
derived using $A_4$ and a reduced seesaw mechanism. It has a simple 
solution as shown by Eqs.~(13) and (14).  It is numerically 
consistent with all present data and predicts the exciting 
possibilty of maximal $CP$ violation in the neutrino sector.

This work is supported in part by the U.~S.~Department of Energy 
under Grant No.~DE-AC02-06CH11357.

\newpage
\bibliographystyle{unsrt}

\end{document}